\documentstyle[12pt,aasms4]{aastex}

\newcommand {\ie} {{\it  i.e.}}

\newcommand {\eg} {{\it e.g.}}
\newcommand {\etal} {{\it et~al.}}

\newcommand {\go} {\mathrel{\hbox{\rlap{\lower.55ex \hbox {$\sim$}}
          \kern-.3em \raise.4ex \hbox{$>$}}}}
\newcommand {\lo} {\mathrel{\hbox{\rlap{\lower.55ex \hbox {$\sim$}}
          \kern-.3em \raise.4ex \hbox{$<$}}}}
\def\Msun{\ifmmode M_\odot \else $M_\odot$\fi} 
\def\Mdot{\ifmmode \dot M \else $\dot M$\fi} 
\def\hmxb{\ifmmode n_{\rm HMXB} \else $n_{\rm HMXB}$\fi}
\def\psnb{\ifmmode n_{\rm PSNB} \else $n_{\rm PSNB}$\fi}
\def\lmxb{\ifmmode n_{\rm LMXB} \else $n_{\rm LMXB}$\fi}
\def\mrp{\ifmmode n_{\rm MRP} \else $n_{\rm MRP}$\fi}
\def\tauh{\ifmmode \tau_{\rm HMXB} \else $\tau_{\rm HMXB}$\fi}
\def\taup{\ifmmode \tau_{\rm PSNB} \else $\tau_{\rm PSNB}$\fi}
\def\tauel{\ifmmode \tau_{\rm LMXB} \else $\tau_{\rm LMXB}$\fi}
\def\taum{\ifmmode \tau_{\rm MRP} \else $\tau_{\rm MRP}$\fi}
\def\ratem{\ifmmode {n_{\rm MRP}\over\tau_{\rm MRP}} \else
${n_{\rm MRP}\over\tau_{\rm MRP}}$\fi} 
\def\ratel{\ifmmode {n_{\rm LMXB}\over\tau_{\rm LMXB}} \else
${n_{\rm LMXB}\over\tau_{\rm LMXB}}$\fi} 
\def\LX{\ifmmode L_X \else $L_X$\fi} 
\def\LB{\ifmmode L_B \else $L_B$\fi} 

\begin{document}

\title{X-RAY PROBES OF COSMIC STAR-FORMATION HISTORY}
\author{Pranab Ghosh$^{1,2}$ \& Nicholas E. White$^2$}

\affil{$^1$Tata Institute of Fundamental Research, Bombay 400 005, India}
\affil{$^2$NASA Goddard Space Flight Center, Greenbelt, MD 20771}

\lefthead{Ghosh \& White}
\righthead{Cosmic Star-formation History}

\begin{abstract}
 
We discuss the imprints left by a cosmological evolution of the star 
formation rate (SFR) on the evolution of X-ray luminosities \LX~of normal 
galaxies, using the scheme proposed by White and Ghosh (1998, WG98),
wherein the evolution of \LX~of a galaxy is driven by the evolution of its 
X-ray binary population. As indicated in WG98, the profile of \LX~with 
redshift can both serve as a diagnostic probe of the SFR profile and 
constrain evolutionary models for X-ray binaries. We report here the first 
calculation of the expected evolution of X-ray luminosities of galaxies, 
updating the WG98 work by using a suite of more recently developed SFR 
profiles that span the currently plausible range. The first {\it Chandra\/} 
deep imaging results on \LX -evolution are beginning to probe the SFR 
profile of bright spirals: the early results are consistent with predictions 
based on current SFR models. Using these new SFR profiles, the resolution of 
the ``birthrate problem'' of low-mass X-ray binaries (LMXBs) and recycled, 
millisecond pulsars (WG98) in terms of an evolving global SFR is more 
complete. We discuss the possible impact of the variations in the SFR 
profile of individual galaxies and galaxy-types.
 
\end{abstract}


\section{INTRODUCTION}

The global star-formation rate (SFR) has undergone strong cosmological 
evolution: it was $\sim$ 10 times its present value at $z\approx 1$, had 
a peak value $\sim$ 10--100 times the present one in the redshift range 
$z\sim$ 1.5--3.5, and declined again at high $z$ (Madau \etal~1996; 
Madau, Pozzetti \& Dickinson 1998, M98; Blain, Smail, Ivison 
\& Kneib 1999, B99a; Blain \etal~1999, B99b, and references therein). 
Details of the SFR at high redshifts are still somewhat uncertain, because 
much of the star formation at $2\lo z\lo 5$ may be dust-obscured and so 
missed by optical surveys, but detected readily through the copious 
submillimeter emission from the dust heated by star formation (Hughes 
\etal~1998; Barger \etal~1999).

The X-ray emission of a normal galaxy (\ie, one without an active nucleus) 
is dominated by the integrated emission of the galaxy's X-ray binary 
population (see, \eg, Fabbiano 1995). In WG98 we discussed the effects
of an evolving SFR on the evolution of X-ray binary populations of 
galaxies, and so on that of the X-ray emission from normal galaxies. We 
suggested that X-ray luminosities \LX~of normal galaxies could show
significant evolution (by up to a factor $\sim$ 10), even in the relatively 
nearby redshift range $z\sim$ 0.5--1.0. We also showed that an evolving SFR 
could resolve the ``birthrate problem'' involving LMXB and their descendant 
``millisecond'' radio pulsars (MRP, see Kulkarni \& Narayan 1988; Lorimer 
1995, L95).

The SFR profile used in WG98---the only profile available at the
time---was based on the optical/UV data alone. Over the past three years,
there has been considerable progress in our understanding of cosmic 
star-formation history. In addition, very deep X-ray imaging with $Chandra$ 
is beginning to detect normal galaxies in the redshift range $z\sim$ 
0.5--1.0, so that comparison with observations is becoming possible for 
the first time. In this {\it Letter}, we consider quantitatively the key 
imprints left by SFR evolution  on the \LX -evolution profiles of normal 
galaxies, using the best SFR models currently available. We briefly discuss 
the recent results of Brandt \etal~(2001, Bran01) from the ultradeep 
$Chandra$ imaging of the Hubble Deep Field North (HDF-N). A detailed 
calculation of the expected X-ray flux distribution of the HDF-N galaxies 
based on the results of this paper is reported by Ptak \etal~(2001, Ptak01). 
We also reconsider the resolution of the LMXB--MRP birthrate problem using 
the new SFR profiles, and find that the calculated rates for both (a) whole 
populations of LMXB and MRP, and, (b) short-period systems, are consistent 
with observation for some SFR profiles suggested recently to account for 
the multiwaveband SFR data. We discuss the relative roles of global SFR 
profiles on the one hand, and the profiles of individual galaxies or 
galaxy-types on the other.

\section{X-RAY LUMINOSITY EVOLUTION WITH EVOLVING SFR}

The total X-ray output of a normal galaxy can be modeled as the sum 
of those of its HMXB and LMXB as per the WG98 scheme, wherein the 
evolution of each species ``$i$'' is described by a timescale $\tau_i$ 
(see WG98). To study the effects of the dependence of $\tau_i$ on the 
binary period and other evolutionary parameters, we run the evolutionary 
scheme over ranges of likely values of $\tau_i$ given in the literature. 
The evolution of the HMXB population in response to an evolving 
star-formation rate SFR(t) is given by
\begin{equation}
{\partial\hmxb(t)\over\partial t} = \alpha_h {\rm SFR}(t)-{\hmxb(t)
\over\tauh},
\label{eq:evohmxb}
\end {equation}
where \hmxb~is the number density of HMXBs in the galaxy, and
\tauh~is the HMXB evolution timescale. $\alpha_h$ is the rate of 
formation of HMXBs per unit SFR, given approximately by $\alpha_h = 
{1\over 2}f_{\rm binary}f^h_{\rm prim}f^h_{\rm SN}$, where $f_{\rm 
binary}$ is the fraction of all stars in binaries, $f^h_{\rm prim}$ is 
that fraction of primordial binaries which has the correct range of 
stellar masses and orbital periods for producing HMXBs (van den Heuvel 
1992, vdH92 and the references therein), and $f^h_{\rm SN}\approx 1$ is 
that fraction of massive binaries which survives the first supernova. 
In these calculations, we have adopted a representative value $\tauh
\sim 5\times 10^6$ yr according to current evolutionary models. In our 
introductory model here, \tauh~includes both (a) the time taken ($\sim 
4-6\times 10^6$ yr) by the massive companion  of the neutron star to
evolve from the instant of the neutron-star-producing supernova to the 
instant when the ``standard'' HMXB phase begins, and, (b) the duration 
($\sim 2.5\times 10^4$ yr) of this HMXB phase (vdH92 and references 
therein). Since the second timescale is negligible compared to the first, 
little error is made by approximating this two-step process by a single 
step with an overall timescale \tauh.

Two basic methods of LMXB production have been discussed. In the cores 
of dense globular clusters, they can be produced by the tidal capture 
of a neutron star by a normal star. Over the rest of a galaxy, stellar 
densities are insufficient for tidal capture, and LMXBs are produced 
by the evolution of primordial binaries (see, \eg, Webbink, Rappaport 
\& Savonije 1983; Webbink 1992). In this paper, we consider only the 
latter mechanism. For spiral galaxies, at least, this must be the 
dominant mechanism, since the globular-cluster LMXB population in such 
galaxies only accounts for a relatively small fraction of the total 
X-ray luminosity.

LMXB evolution from primordial binaries has two distinct stages (WG98) 
after the supernova produces a post-supernova binary (PSNB) containing
the neutron star. First, the PSNB evolves on a timescale \taup~due to 
nuclear evolution of the neutron star's low-mass companion and/or 
decay of binary orbit due to gravitational radiation and magnetic 
braking, until the companion comes into Roche lobe contact and the LMXB 
turns on. Subsequently, the LMXB evolves on a timescale \tauel. Since 
\taup~and \tauel~are comparable in general, we must describe the two 
stages separately (WG98) by:
\begin{equation}
{\partial\psnb(t)\over\partial t} = \alpha_l {\rm SFR}(t)
-{\psnb(t)\over\taup},
\label{eq:evopsnb}
\end{equation}
\begin{equation}
{\partial\lmxb(t)\over\partial t} = {\psnb(t)\over\taup}
-{\lmxb(t)\over\tauel},
\label{eq:evolmxb}
\end{equation}
Here, \psnb~and \lmxb~are the respective number densities of PSNB and
LMXB in the galaxy, and $\alpha_l$ is the rate of formation of LMXB per 
unit SFR, given approximately by $\alpha_l = {1\over 2}f_{\rm binary}
f^l_{\rm prim}f^l_{\rm SN}$, where $f_{\rm binary}$ is the fraction of 
all stars in binaries, $f^l_{\rm prim}$ is that fraction of primordial 
binaries which has the correct range of stellar masses and orbital 
periods for producing LMXBs, and $f^l_{SN}$ is that fraction of such 
binaries which survives the massive star's supernova.

We display evolution in terms of the redshift $z$, which is related to
the cosmic time $t$ by $t_9 = 13(z+1)^{-3/2}$, where $t_9$ is $t$ in
units of $10^9$ yr, and a value of $H_0=50$ km s$^{-1}$ Mpc$^{-1}$ has
been used\footnote{For ease of comparison with WG98 and M98, we continue
to use here a Friedman cosmology with $q_0 = 1/2$; other cosmologies will
be considered elsewhere. Other values of the Hubble constant lead to a
straightforward scaling: for $H_0=70$ km s$^{-1}$ Mpc$^{-1}$, for 
example, $t_9 \approx 10(z+1)^{-3/2}$, so that the results remain 
unchanged if all timescales are shortened by a factor of 1.3.}. 
We consider the suite of current SFR models detailed in Table 1 to cover 
a plausible range, using the parameterization ofB99a,b. Models of the 
``peak'' class have the form
\begin{equation}
{\rm SFR}_{\rm peak}(z) = 2\left(1+\exp{z\over z_{max}}
\right)^{-1}(1+z)^{p+{1\over 2z_{max}}},
\label{eq:SFRpeak}
\end{equation}
while those of the ``anvil'' class have the form
\begin{equation}
{\rm SFR}_{\rm anvil}(z) = \cases{(1+z)^p,&{$z\leq z_{max}$},\cr
                                  (1+z_{max})^p,&{$z> z_{max}$}.\cr}
\label{eq:SFRanvil}
\end{equation}
These functional forms are not unique, but useful, since they have a
convenient low-$z$ limit, SFR$(z)\propto (1+z)^p$, where all SFR profiles 
must agree with the  optical/UV data (M98), and since the model
parameters can be manipulated to mimic a wide range of star-formation
histories (B99b). Peak-class profiles are useful for describing (a) SFRs
determined from optical/UV observations, \ie, Madau-type (M98) profiles,
called ``Peak-M'' in Table 1, and, (b) more general SFRs with enhanced 
star formation at high $z$, a good example of which is the 
``hierarchical'' model of B99b, wherein the submillimeter emission is 
associated with galaxy mergers in an hierarchical clustering model of 
galaxy evolution. Anvil-class profiles are useful for describing the 
results of ``monolithic'' models. The ``Gaussian'' model (B99a,b)is an 
attempt at giving a good account of the SFR at both low and high $z$ by 
making a composite of the Peak-G model (see Table 1) and a Gaussian
starburst at a high redshift $z_p$, \ie, a component
\begin {equation}
{\rm SFR}_{\rm Gauss}(z) = \Theta\exp\left\{-{[t(z)-t(z_p)]^2\over
2\sigma^2}\right\}.
\label{eq:SFRGauss}
\end{equation}
Based on the $IRAS$ luminosity function, this component is devised to 
account for the high-$z$ data, particularly the submillimeter 
observations (B99a). For its parameters (see Table 1), we have used the 
revised values given in B99b. In all models considered here, no galaxies 
exist for sufficiently large redshifts, $z>$ 10.

Figures 1 and 2 show the prompt evolution of HMXBs and the slow evolution 
of LMXBs, and the evolution of the total X-ray binary population, where 
the two components have been so weighted as to represent the total X-ray 
emission from the galaxy (the weight-ratio is the product of $\alpha_h/
\alpha_l$ and the ratio of the average luminosity of HMXBs and LMXBs). 
The HMXB profile closely follows the SFR profile because \tauh~is small 
compared to the SFR evolution timescale. By contrast, the LMXB profile has 
a significant lag behind the SFR profile because \taup~and \tauel~are 
comparable to SFR evolution timescale: depending on these timescales, the 
LMXB profile peaks at redshifts $\sim$ 1--3 later than the HMXB profile 
(as seen clearly in Figs. 1 and 2), which is a characteristic signature 
of SFR evolution (WG98). 

Figure 1 compares the \LX -evolution corresponding to the (Madau or Peak-M) 
SFR profile and the evolutionary timescales we originally used in WG98, \ie,
(a) \taup~= 1.9 Gyr, \tauel~= 0.1 Gyr for the whole LMXB population, and
(b) \taup~= 0.9 Gyr, \tauel~= 0.5 Gyr for the short-period systems. In
Figure 2, we display the \LX -evolution for a range of SFR
profiles---Peak-M, Hierarchical, Anvil-10, and Gaussian, the evolutionary
timescales being held fixed at \taup~= 1.9 Gyr, \tauel~= 1.0 Gyr. Between
them, the two figures thus explore the effects of (a) varying the
evolutionary timescales for a fixed SFR profile, {\it viz.\/} Peak-M, and
(b) varying the profile for a fixed set of evolutionary timescales. For 
sufficiently {\it slow\/} LMXB evolution, the galaxy's X-ray emission is 
dominated by LMXBs at low redshifts ($0\lo z\lo 1$), and by HMXBs at high 
redshifts. As a result, the total \LX -profile is {\it strongly\/} 
influenced at high redshifts by the SFR profile (see Fig.2), and may 
actually show hints of a double-peak structure (WG98) at intermediate 
redshifts (\eg, the Gaussian profile in Fig.2). For sufficiently 
{\it fast\/} LMXB evolution, on the other hand, LMXBs and HMXBs may have 
comparable contributions to \LX~at low redshifts (see Fig.1).     

Bran01 estimate that the average X-ray luminosity of the bright spiral 
galaxies at an average redshift $z\approx 0.5$ used in their stacking 
analysis is about a factor of 3 higher than that in the local Universe 
($z<0.01$). This observed evolution, \LX (0.5)/\LX (0.0) $\sim$ 3, can be 
compared with our theoretical results in Table 2. The degree of evolution 
from $z=0$ to $z=$ 0.5--1.0 increases from Madau-type profiles to those with 
additional star formation at high redshifts, the numbers for the Peak-M 
profile being in best agreement with Bran01. For a given profile, the 
evolutionary factor is smaller for a slower evolution of LMXBs (roughly
measured in this context by the total time \taup~+ \tauel), as expected. 
More sophisticated estimates of the expected \LX -distribution of HDF-N 
galaxies, based on these evolutionary scenarios, are described in Ptak01.

\section{BIRTHRATE PROBLEM: FURTHER CONSIDERATIONS}

As discussed in WG98, the evolution of MRPs from LMXBs is described by an
equation similar to Eq.~3, but involving the MRP number density \mrp, and 
MRP evolution timescale $\taum\sim 3\times 10^9 - 3\times 10^{10}$ yr 
(Camilo \etal~1994). Our evolutionary scheme yields the profiles of LMXB 
and MRP evolution for a given SFR profile, giving the number ratio, $N_r
\equiv \mrp/\lmxb$, and  the rate ratio, $R_r\equiv\ratem/\ratel$, of the 
MRP and LMXB populations at the present epoch ($z=0$). We showed in WG98 
that the Peak-M profile yielded $R_r\simeq 1$ for the overall MRP and LMXB 
populations, in agreement with current observations (L95). However, for 
short-period systems (LMXB periods $\lo 3$ days), this profile yielded a 
value $R_r\simeq 3$, smaller than the current value $R_r\approx 8$ 
estimated from observations (L95). On repeating these calculations with 
other SFR profiles which have additional star formations at high 
redshifts, \eg, the Gaussian and hierarchical profiles, we now find 
reasonable agreement for both whole populations and short-period systems,
since these give $R_r\approx 6-8$ for short-period systems, for plausible 
evolutionary timescales. Thus, there appears to be no discrepancy between
observation and the idea that SFR evolution naturally leads to values of
$R_r$ which can be $\approx$ 1 as well as considerably above unity (but
see \S 4).

A new development in SFR research since WG98 has been the study of 
star-formation histories of individual galaxies and various galaxy-types. 
SFR profiles of individual galaxies, ranging from those in the Local Group 
to those in the HDF at redshifts $0.4\lo z\lo 1$, have been inferred, 
using a variety of techniques (Glazebrook \etal~1999, Abraham \etal~2000, 
Hernandez \etal~2000). For various galaxy-types, models of 
Spectro-photometric evolution, which use the synthesis code 
{\it P\'{e}gase\/} and are constrained by deep galaxy counts, have been
developed (Rocca-Volmerange and Fioc 2000), leading to a model SFR profile
for each type. The birthrate problem was originally formulated entirely in 
terms of observations in our own galaxy, so that, strictly speaking, we 
should use the SFR profile of our galaxy. As none of the above techniques 
can be used in our own galaxy, it is difficult (but not impossible) to 
determine its SFR profile: calculations with such a profile will be 
described elsewhere. 
 
\section{DISCUSSION}

We have shown that different global SFR profiles within the currently
admissible range lead to very different \LX -evolution profiles, so that
the latter profiles can be an {\it independent\/} X-ray probe of cosmic 
star-formation history. Our results and a more detailed consideration by 
Ptak01 indicate that the early {\it Chandra\/} results (Bran01) are 
consistent with the Peak-M profile: let us now clarify the underpinnings 
of this result by applying the considerations of global and individual 
SFR profiles summarized above. Bran01 used bright spirals for their 
stacking analysis. Rocca-Volmerange and Fioc (2000) have shown that the 
model SFR profile for such (Sa-Sbc) spirals rises roughly in a Madau 
fashion from $z=0$ to $z\approx 1$ (which these authors ascribe to a bias 
in the original sample used to construct the Madau profile towards bright 
spirals), and thereafter flattens to a roughly constant value $\sim$ 12 
times that at $z=0$, falling again at $z\go 7$. We can roughly represent 
this profile in the range $0<z\lo 7$ by an anvil-type profile (see \S2),
with the parameter $z_{max}$ as given in Table 1, and the parameter
$p\approx$ 2.7. For such a profile with the timescales \taup~= 1.9 Gyr, 
\tauel~= 1.0 Gyr, as in Figure 2, our evolutionary scheme gives  
\LX (0.5)/\LX (0.0) = 3.3, and \LX (1.0)/\LX (0.0) = 5.4, in good 
agreement with both the Bran01 results and the Peak-M results given in
Table 2. We now see why the Peak-M profile appears to give a good account 
of the Bran01 results. In effect, the Bran01 analysis may be probing the 
SFR profile of only the bright spirals in HDF-N, and the fact that the 
Peak-M profile is consistent with the Bran01 results does {\it not\/} 
imply that the global SFR necessarily follows the Peak-M profile. Thus, 
there is no basic contradiction between the results of \S2 and \S3. 
However, the fact that our global analysis of the birthrate problem (\S3) 
gives the best agreement with observation for SFR profiles (Gaussian and 
hierarchical) which have larger SFRs at high redshifts than the model 
Sa-Sbc profile of Rocca-Volmerange and Fioc (which, in turn, has a larger 
SFR at high redshifts than the Peak-M profile) remains to be understood 
fully. The individual-galaxy treatment may hold the key here.           

Signatures of star formation in the relatively nearby ($z\lo 1$) Universe 
have usually been sought at wavelengths other than X-rays. In this redshift
range, the X-ray probe described here is a ``fossil record'' study of star 
formation at earlier epochs for sufficiently slow LMXB evolution (see
Fig.2), since LMXBs then preserve this record for $\sim$ 1 Gyr after the 
star-formation peak is gone. Conversely, for a known SFR profile, this is a 
unique proving ground for theories of LMXB evolution, which have been built 
almost wholly on the basis of our experience in the current epoch: this new 
probe is, at present, our only direct means of watching LMXB evolution 
unfold over cosmic time.   
 
\acknowledgements

It is a pleasure to thank L. Angelini, R. Griffiths, R. Mushotzky,
and A. Ptak for stimulating discussions.

{}


\begin{table}

\caption{Star Formation Rate (SFR) Profiles$^a$}
\halign{%
\rm#\hfil&\qquad\rm#\hfil&\qquad\rm#\hfil
&\qquad\rm#\hfil&\qquad\hfil\rm#\qquad\hfil\cr
\noalign{\vskip6pt}\noalign{\hrule}\noalign{\vskip2pt}
\noalign{\hrule}\noalign{\vskip6pt}
Model&$z_{max}$&$p$&Comments\cr
\noalign{\vskip6pt}\noalign{\hrule}\noalign{\vskip6pt}
Peak-M&0.39&4.6&Madau profile\cr
Hierarchical&0.73&4.8&Hierarchical clustering model$^b$\cr
Anvil-10&1.49&3.8&Monolithic models\cr
Peak-G&0.63&3.9&``Peak'' part of composite Gaussian\cr
Gaussian&N/A&N/A&Gaussian starburst$^c$ added at high-$z$\cr
\noalign{\vskip6pt}\noalign{\hrule}\noalign{\vskip12pt}}
$^a$Model parameters taken from B99a,b.
\vskip3pt
$^b$We have chosen the B99b model with a dust temperature 45 K. Note that
the parameter $p$ for the SFR equals 3/2 plus the value of the parameter
$p$ occurring in equation (16) for the merger efficiency in B99b.
\vskip3pt
$^c$Parameters of Gaussian starburst component (see eq.[6] of text), are
from the modified model given in B99b, \ie, $z_p = 1.7$, $\sigma = 1.0$
Gyr, and $\Theta = 70$.

\end{table}

\begin{table}

\caption{Evolution of X-ray Luminosity \LX}
\halign{%
\rm#\hfil&\qquad\rm#\hfil&\qquad\rm#\hfil
&\qquad\rm#\hfil&\qquad\hfil\rm#\qquad\hfil\cr
\noalign{\vskip6pt}\noalign{\hrule}\noalign{\vskip2pt}
\noalign{\hrule}\noalign{\vskip6pt}
Model&$\taup$&$\tauel$&${\LX(0.5)\over\LX(0.0)}$&
${\LX(1.0)\over\LX(0.0)}$\cr
\noalign{\vskip6pt}\noalign{\hrule}\noalign{\vskip6pt}
Peak-M&1.9&0.1&3.9&5.4\cr
Peak-M&0.9&0.5&4.6&6.8\cr
Peak-M&1.9&1.0&3.4&4.1\cr
Hierarchical&1.9&1.0&6.2&11.3\cr
Anvil-10&1.9&1.0&5.4&10.1\cr
Gaussian&1.9&1.0&7.5&16.0\cr
\noalign{\vskip6pt}\noalign{\hrule}\noalign{\vskip12pt}}

\end{table}

\vfill
\eject

\begin{figure}
\figurenum{1}
\centerline{
\includegraphics[scale=1.0]{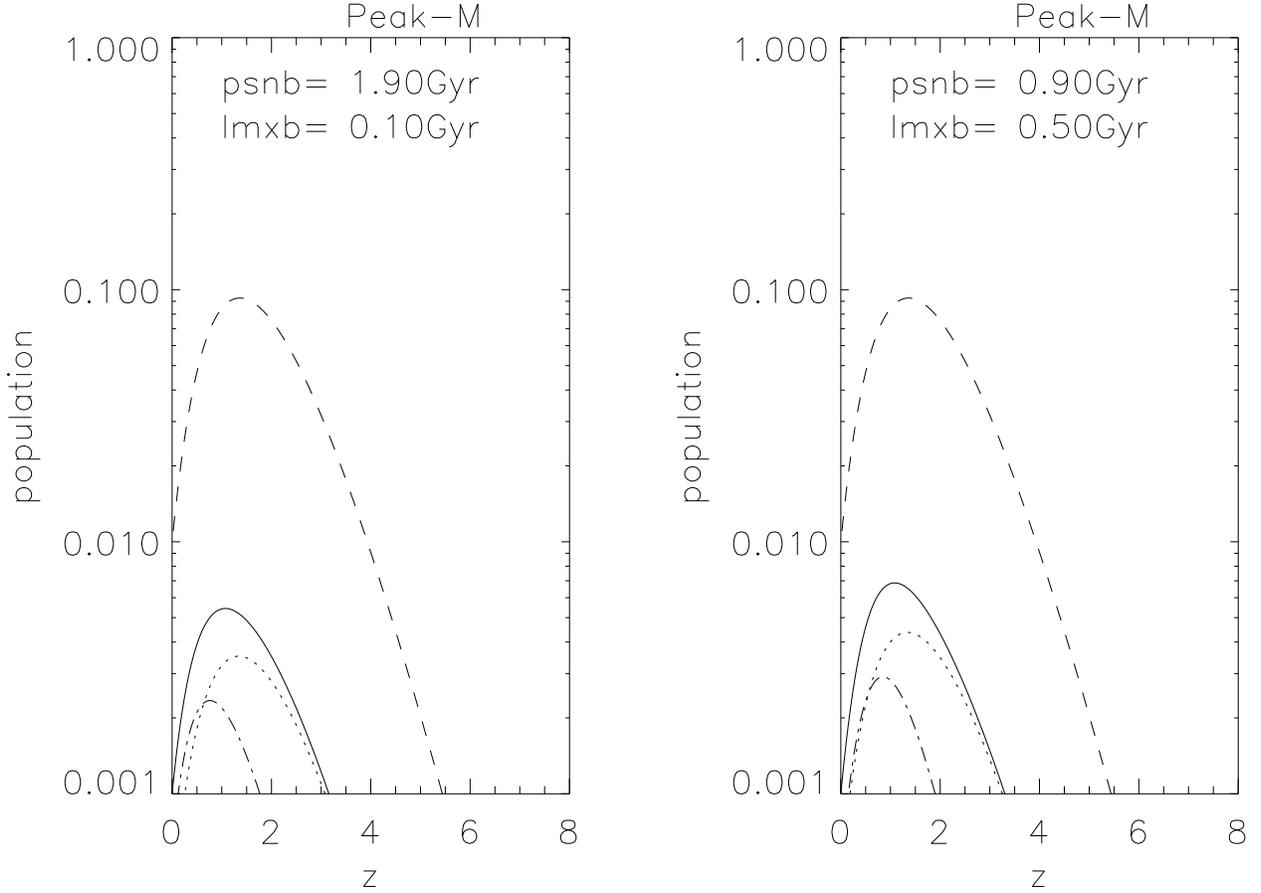}}

\caption{Evolution of HMXB population (dotted line), LMXB population
(dash-dotted line), and the total X-ray luminosity \LX~(solid line) of a
galaxy with a given SFR (dashed line). As absolute ordinate scales are
irrelevant for these evolutionary profiles, they have been adjusted for
convenience of display: \LX~always starts at 0.001 at $z=0$, so that its
evolution can be immediately read off the figure, and SFR always starts at
0.01 at $z=0$, so that different SFR profiles can be readily compared.
This figure is for the Peak-M profile (see text), showing the effect of
varying the evolutionary timescales \taup~and \tauel, whose values are
written on each panel. The timescales used here are those used in WG98,
corresponding to whole LMXB populations and short-period systems.}

\end{figure}

\begin{figure}
\figurenum{2}
\centerline{
\includegraphics[scale=1.0]{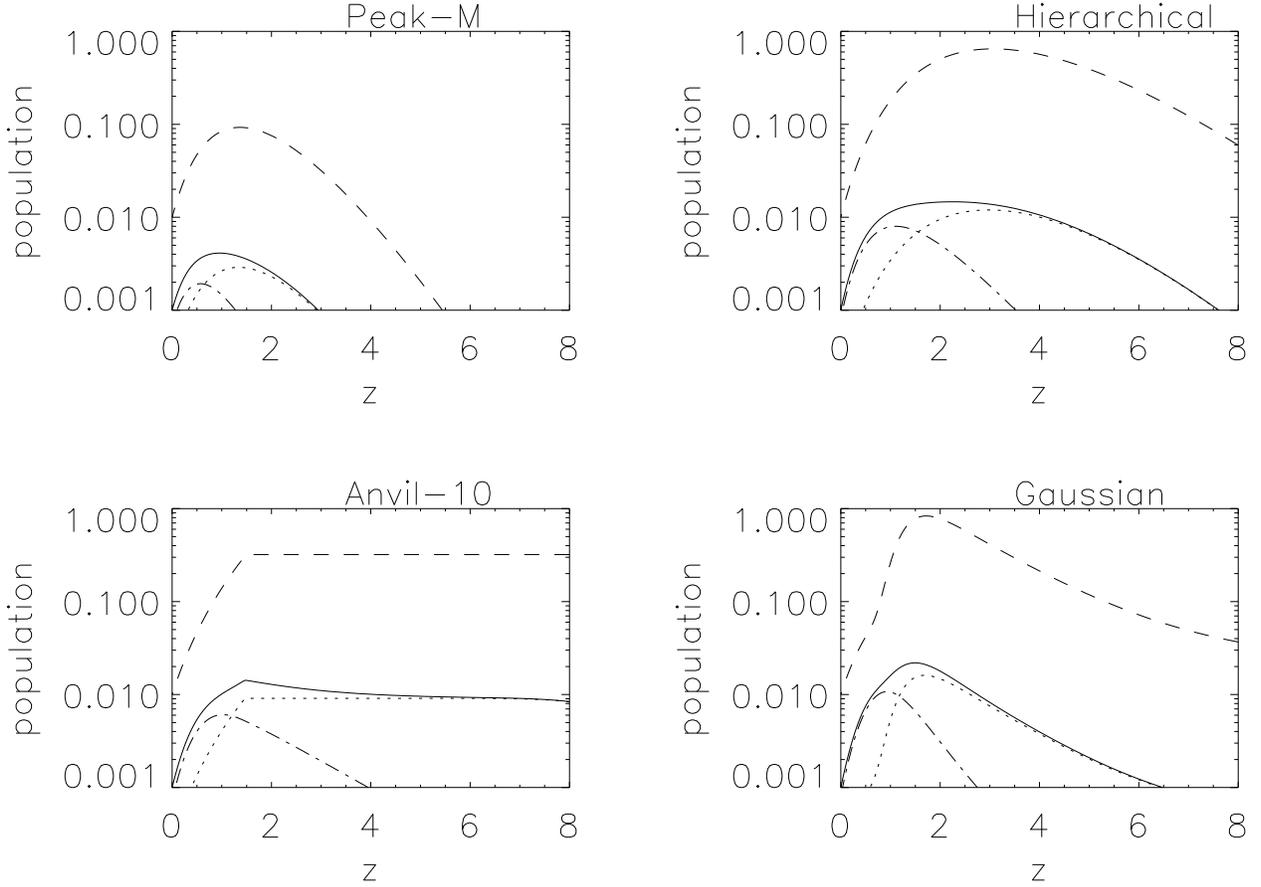}}

\caption{Same as Figure 1, but showing the effect of varying the SFR
profile. The evolutionary timescales are kept fixed at \taup~= 1.9 Gyr
and \tauel~= 1.0 Gyr for all cases, and SFR profiles from Table 1 are used.
Each panel is labeled by the name of its SFR profile. Evolutionary factors
from this and the previous figure are collected in Table 2 and described
in the text.}

\end{figure}

\end{document}